# Localized Resonant Phonon Polaritons in Biaxial Nanoparticles

*Daniel Beitner[1,2,3,\*], Asaf Farhi[3,\*], Ravindra Kumar Nitharwal[4], Tejendra Dixit[5], Tzvia Beitner[1], Shachar Richter[1,2], SivaRama Krishnan[4] and Haim Suchowski[2,3]*

\* D. Beitner and A. Farhi contributed equally to this work

[1]*Department of Materials Science and Engineering Faculty of Engineering Tel Aviv, University Ramat Aviv, Tel Aviv 69998, Israel*

[2]*University Centre for Nanoscience and Nanotechnology Tel Aviv University Ramat Aviv, Tel Aviv 69998, Israel*

[3]*School of Physics and Astronomy, Faculty of Exact Sciences, Tel Aviv University, Tel Aviv 69978, Israel*

[4]*Department of Physics, Indian Institute of Technology, Madras, Chennai 600036, India*

[5]*Optoelectronics and Quantum Devices Group, Department of Electronics and Communication Engineering, Indian Institute of Information Technology Design and Manufacturing (IIITDM) Kancheepuram*

**Abstract**
The discovery of localized plasmon polariton resonances has been pivotal in enabling tunability of the optical resonance. Recently, extensive research efforts have aimed to expand these achievements to other polaritonic states that exhibit less loss and in other spectral regions. However, these efforts were limited to isotropic or uniaxial structures, and an eigenmode theory was derived only for isotropic particles. Here, we present a breakthrough in synthesizing biaxial nanostructures that exhibit localized hyperbolic phonon resonances with high Q-factors in the mid-infrared. Furthermore, we develop a theory that predicts high-order resonances in anisotropic particles with coupling between the axial permittivites. Finally, we confirm the theoretical predictions through near-field measurements, which demonstrate the existence of both the first and higher-order resonant modes. Our findings provide the foundation for designing a new generation of anisotropic resonators with various applications in the mid-IR range. Our analysis applies to other fields, such as quasi-magnetostatics and heat conduction.

## Introduction

When optical phonons, the out-of-phase lattice vibrational modes in polar crystals, are coupled to photons, they give rise to quasiparticles known as phonon polaritons (PhPs). These PhPs have been heavily studied in polar dielectrics[1], polar semiconductors, and recently in two-dimensional (2D) materials[2,3]. The potential of PhPs to manipulate light at the nanoscale, especially in the coveted infrared (IR) spectral range, is critical for the development of new IR-active metamaterials technologies.

Hyperbolic PhPs (HPhPs) occur in extremely anisotropic materials when the real component of the permittivity is in opposite signs along different directions[4,5]. They have recently garnered considerable attention due to their tightly confined and directional nature. Such characteristics make HPhPs highly advantageous for various optical applications, including enhanced spontaneous emission[6,7], hyper-lensing[8–12], high-sensitivity chemical sensing [13,14], negative refraction[15–17], and waveguiding[18,19].

Recently, several naturally anisotropic polar materials capable of supporting HPhPs, such as hexagonal boron nitride (h-BN)[3,20–22], α-phase molybdenum trioxide (α-$MoO_3$)[23], and α-phase vanadium pentoxide [24], have been identified. Among these, α-$MoO_3$ has gained notable attention due to its biaxial anisotropy with different optical indices in each axis. This property leads to tunable and long-lived HPhPs active in the atmospheric mid-infrared (mid-IR) window. To date, HPhPs in α-$MoO_3$ have been demonstrated in flakes[25–28], macro-disks[23,29], nanobelts[5,30], nanocavities[31], and metamaterials[32]. Furthermore, diverse strategies for modifying HPhPs in α-$MoO_3$ have been explored, including adjustments to layer thickness, inter-layer twist angle, coupling with plasmonic nanoparticles, alterations in the dielectric environment, and geometric confinement. While recent works on sub-wavelength geometries have shown the existence of localized hyperbolic phonon resonances (LHPhR) in uniaxial materials such as hBN for nanodisks[33], nanobars[34], and nano pillars[22], the research on HPhPs in biaxial materials has only centered on *propagating* HPhPs modes on 2D α-$MoO_3$ surfaces or large macro-sized particles.

Here, we present the first experimental observation of LHPhR in sub-wavelength biaxial nanoparticles. The nanoparticles, made of α-$MoO_3$, were synthesized using femtosecond pulsed laser ablation in a liquid (fs-PLAL) and characterized spatially and spectrally using a mid-IR scattering scanning near-field microscope (s-SNOM). Our findings reveal that α-$MoO_3$ nanoparticles have a diverse optical spectrum of phononic polaritons spanning the 900-1000 $cm^{-1}$ range. The wide range of possible aspect ratios for these nanoparticles allows for broad tunability of the resonances of the phononic polaritons. In addition, multiple modes, including high-order directional polaritons, are observed.

We also develop a novel theoretical model describing biaxial nanoparticle modes and eigen-permittivity relations to back up our observations. In addition to numerical simulations of these nanoparticles, this model sheds light on the nature of these photonic modes, offering insights into the design of tailor-made resonances in anisotropic particles. Specifically, we predict and demonstrate that biaxial particles support three distinctly oriented LHPhR modes, with coupling between the different crystal axes in the high-order modes.

Our study introduces a new class of mid-IR nanoparticle resonators exhibiting extremely large quality factors with controlled directionality, which could usher in a new era of detectors, image sensors, and highly sensitive photonic devices in the mid-IR range.

## Results:

α-MoO$_3$ nanoparticles were synthesized using fs-PLAL, which enables the production of particles of varying sizes. Figure 1a shows a schematic of the process. A slab of the material is placed in water, and an fs laser source is focused on the surface. The laser pulses ablated the material to create nanoparticles. After ablation, the particles are dispersed in deionized (DI) water, cleaned, and subsequently deposited onto a Si substrate through drop casting. X-ray diffraction (XRD) and Raman spectra taken from the ablated particles confirm their composition as α-MoO$_3$, as shown in Figure 1b and the supplementary information (SI) Figure S1.

α-MoO$_3$ is a biaxial hyperbolic material with three distinct Reststrahlen bands (RB) in each crystal direction of its unit cell. Within the spectral range of the RB band, the real dielectric value becomes negative in that crystal direction. Specifically, the lowest energy band, RB1 (545-851 cm$^{-1}$), extends along the [001] direction, RB2 (822–962 cm$^{-1}$) extends in the [100] direction, and RB3 (957–1007 cm$^{-1}$) is aligned with the [010] direction as seen in Figure 1c. The ablated α-MoO$_3$ particles tend to take on the form of either small oblate nano-ellipsoids with a small axis of diameter of 50-120 nm and two equal large axes of diameter of 180-220 nm or nanorods with a thickness of approximately 120 nm and lengths ranging from 200 to 1000 nm, as seen in the schematic in Figure 1d and the SI Figure S1. The orientation directions were determined using a combination of the measured near-field spectra and observed resonance shape as compared to the analytical model of the system, see Figure 1d.

The s-SNOM measurement system used in this study is illustrated in Figure 1d. A tunable continuous wave quantum cascade laser (QCL) is used in conjunction with a pseudo-heterodyne detection scheme. This configuration allows for recording the complex optical signal denoted by $\sigma_n = S_n \cdot e^{\phi_n i}$, where $\sigma_n$, $S_n$ and $\phi_n$ are the $n^{th}$ order harmonic of the s-SNOM total signal, amplitude, and phase, respectively. We used the s-SNOM to spatially image the different resonance modes and record optical phase and amplitude spectra in the 850-1050 cm$^{-1}$ (8.5-11 um) spectral range.

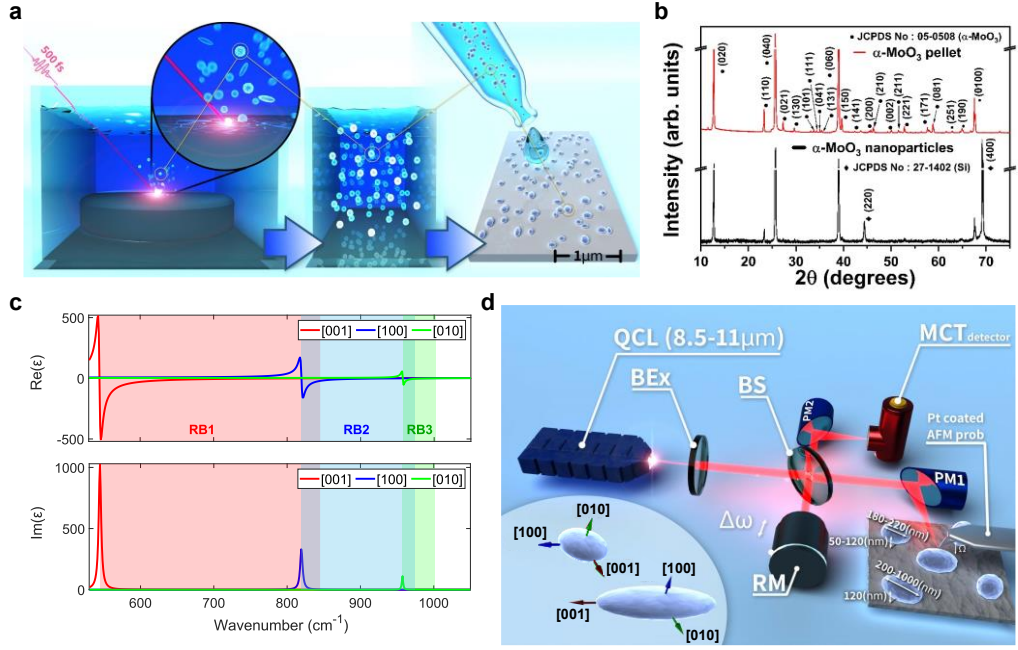

*Figure 1 – fs-PLAL process to create α-MoO₃ nanoparticles and their characterization. a) – Schematic of the fs-PLAL process: In the first stage, the fs laser is focused on the surface of a slab of α-MoO₃, causing the ablation of nanoparticles. The nanoparticles are washed in DI water, and particles are drop-cast on Si surface. b) XRD pattern of α-MoO₃ nanoparticles (black line) and α-MoO₃ pellet (red line). c) Calculated real and imaginary parts of the dielectric permittivity of α-MoO₃ along the three crystal directions ([010], [001], and [100])[35]. The shaded regions indicate the area of the three Reststrahlen bands. b) Size distribution of α-MoO₃ nanoparticles created by fs-PLAL measured via AFM. d) Simplified s-SNOM system schematic with components Quantum Cascade Laser (QCL), Beam Expander (BEx), Beam splitter (BS), Parabolic Mirror one and two (PM), Reference Mirror (RM) oscillating at Δω, Pt coated AFM probe ) oscillating at Ω, and HgCdTe (MCT) liquid N₂ cooled amplified photodetector. The different studied α-MoO₃ nanoparticles and their dimensions are shown along with the presumed crystal directions from measurements.*

A complete theory for the optical response of anisotropic nanoparticles can be described as follows: particles with a length scale much smaller than the wavelength of the incoming field can be analyzed in the quasistatic regime, where they support localized standing modes. To calculate the eigenstates of LHPhR in α-MoO₃ nanoparticles, we solve Laplace's equation [36,37] for an anisotropic media without a source $\nabla \cdot \overleftrightarrow{\epsilon} \nabla \psi_n = 0$, where $\psi_n$ is an eigenstate and $\overleftrightarrow{\epsilon}$ is the permittivity tensor, $\overleftrightarrow{\epsilon_1}$ is the inclusion of eigen-permittivity and $\epsilon_2 = 1$ is the host medium permittivity[38]. We can express the dipole-mode eigenstates and eigen-permittivities for a sphere and ellipsoid as (see SI 2 for full details): [39,40]

$$\tilde{\psi}_{sphere,l=1,m=0} = \frac{1}{\sqrt{a}} \begin{cases} \frac{r}{a}\cos\theta & r < a \\ \frac{a^2}{r^2}\cos\theta & r \geq a \end{cases}, \quad \epsilon_{1z} = -2, \quad (1)$$

$$\tilde{\psi}_{l=1,m=0,\text{in el}} \propto z, E_{out\ el} = E_1\left[e_z + \left(\frac{\epsilon_{1z}}{\epsilon_2} - 1\right)\frac{z}{c}\left(\frac{x}{a},\frac{y}{b},\frac{z}{c}\right)\right], \epsilon_{1z} = \epsilon_2 \frac{(L_z - 1)}{L_z},$$

where $L_z = \frac{abc}{2}\int_0^\infty \frac{dq}{(c^2+q)[R(u)]^{1/2}}$, $R(u) = (q+a^2)(q+b^2)(q+c^2)$ and the ellipsoid's dimensions $a, b$ and $c$ denote the half length of the ellipsoid in the $x, y$, and $z$ axes, respectively. These 1ˢᵗ order eigenstates and eigen-permittivities are identical to the isotropic modes. In contrast, they can only be oriented along the resonant crystal direction of the α-MoO₃ crystal, regardless of the direction of the polarization of the source as long as it can excite the mode (has a polarization component along the relevant axis).

We analyzed dipole resonances of an anisotropic ellipsoid inclusion [40] and observed that an elongated (shortened) axis has a lower (higher) eigen-permittivity value compared to a sphere, and since the material permittivity is monotonically increasing in these ranges, the resonance should redshift (blueshift). Material permittivity values with small real values have large imaginary parts (see Figure 1c), which weaken the resonance since the eigen-permittivity is usually real [36].

Next, we analyze the high-order anisotropic sphere modes and eigen-permittivities. Our goal is to express the eigenstates in Cartesian coordinates due to the crystal symmetry and obtain relations that satisfy both boundary conditions and Laplace's equation, which are relevant to an anisotropic medium i.e., ones that enable $\epsilon_i \neq \epsilon_j$. Since the field outside the sphere can be expanded in spherical harmonics, the high-order spherical harmonics are suitable only for an isotropic medium, and there is a continuity of the potential on the particle envelope, we conclude that the anisotropic modes are combinations of spherical harmonics.

We find the following second-order anisotropic modes for uniaxial and biaxial spheres, respectively, along with eigen-permittivity relations:

$$\widetilde{\psi}_2^1 = \psi_2^0 + \frac{(\psi_2^2 + \psi_2^{-2})}{2} \propto z^2 - y^2, \quad \widetilde{\psi}_2^2 = \frac{(\psi_2^1 - \psi_2^{-1})}{2} \propto xz,$$
$$E_{2\ sp\ out}^2 = z\boldsymbol{e}_x + x\boldsymbol{e}_z - \frac{5}{a^2}xz(x\boldsymbol{e}_x + y\boldsymbol{e}_y + z\boldsymbol{e}_z), \epsilon_{1y} = \epsilon_{1z} = -1.5,\ \epsilon_{1x} + \epsilon_{1z} = -3, \quad (2)$$

where $\psi$ denotes spherical harmonic modes, $\widetilde{\psi}$ denotes anisotropic modes, and the first transitions are for inside the sphere volume. Interestingly, while the eigen-permittivities are discrete for isotropic inclusions, anisotropic spheres exhibit eigen-permittivity lines for the second mode.

Similarly, for the ellipsoid particle, we find the same modes inside the inclusion volume and the following mode field outside and eigen-permittivity relation:

$$E_{2\ ellipsoid\ out}^2 = z\boldsymbol{e}_x + x\boldsymbol{e}_z + xz\left(\frac{1}{a} + \frac{1}{c}\right)(\epsilon_{1xz\ iso} - 1)\left(\frac{1}{a}x\boldsymbol{e}_x + \frac{1}{b}y\boldsymbol{e}_y + \frac{1}{c}z\boldsymbol{e}_z\right),$$
$$\frac{c\epsilon_{1x} + a\epsilon_{1z}}{a + c} = \epsilon_{1xz\ iso}, \quad \epsilon_{1xz\ iso} = 1 - \left(\frac{a_1 a_2 a_3}{2}(a_x^2 + a_z^2)I_{xz}\right)^{-1}, \quad (3)$$

where $I_{\alpha\beta} \int_0^\infty \frac{du}{(u+a_x^2)(u+a_z^2)R(u)}$ and we have used some of the definitions from Ref.[39]. Figures 2a,b,d, and e show the scattering peaks for the first and second-order modes of ellipsoids with R=(110,110,56) nm and R=(200,60,60) nm. Figures 2c and f show the projections of these modes on the s-SNOM probe when it is aligned along the α-MoO₃ z direction. Interestingly, there is coupling between the permittivities of the different axes, unlike the situation in isotropic particles[36,37]. The resulting mode spectra for the other crystal orientations in nano-ellipsoids and nanorods were also calculated (see SI Figure S2 and S3).

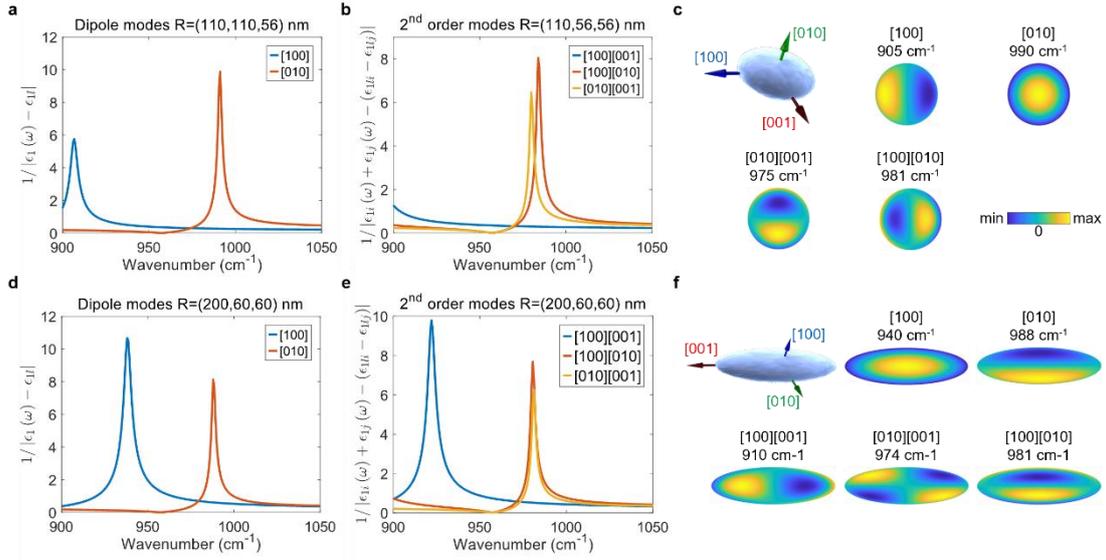

*Figure 2 – α-MoO₃ Eigenstate and electric field projection from analytical model predictions for nano-ellipsoids. a) – b) First and second-order mode scattering peaks of R=(110,110,56) nm³ nano-ellipsoids with the AFM probe aligned along the α-MoO₃ z-[010] for the crystal configuration shown in Figure 1-d upper image. c) 2d (in the x-[100] y-[001] plane) projections of the real part of the electric field in the z-[010] direction. d) – e) First and second-order mode scattering peak of R=(200,60,60) nm³ nano-ellipsoids with the AFM probe aligned along the α-MoO₃ z-[100] for the crystal configuration shown in Figure 1-d lower image. f) 2d (in the x-[001] y-[010] plane) projections of the real part of the electric field in the z-[100] direction.*

Figure 3a presents the normalized near-field 3$^{rd}$ harmonic phase spectra for representative nano-ellipsoids (112/220 nm) over the nanoparticle area. The spectra clearly show a resonance at 988 cm$^{-1}$, with a Q factor of approximately 138. It is worth noting that this measured resonance frequency is blueshifted compared to the bulk α-MoO₃ resonance at approximately 958 cm$^{-1}$ [30]. In addition, the smaller nano-ellipsoids (70/180 nm) that we measured exhibited an even higher Q factor of about 285 (see SI Figure S4). Notably, these values are significantly higher than those previously reported for α-MoO₃ nano-disks and slabs [32,41].

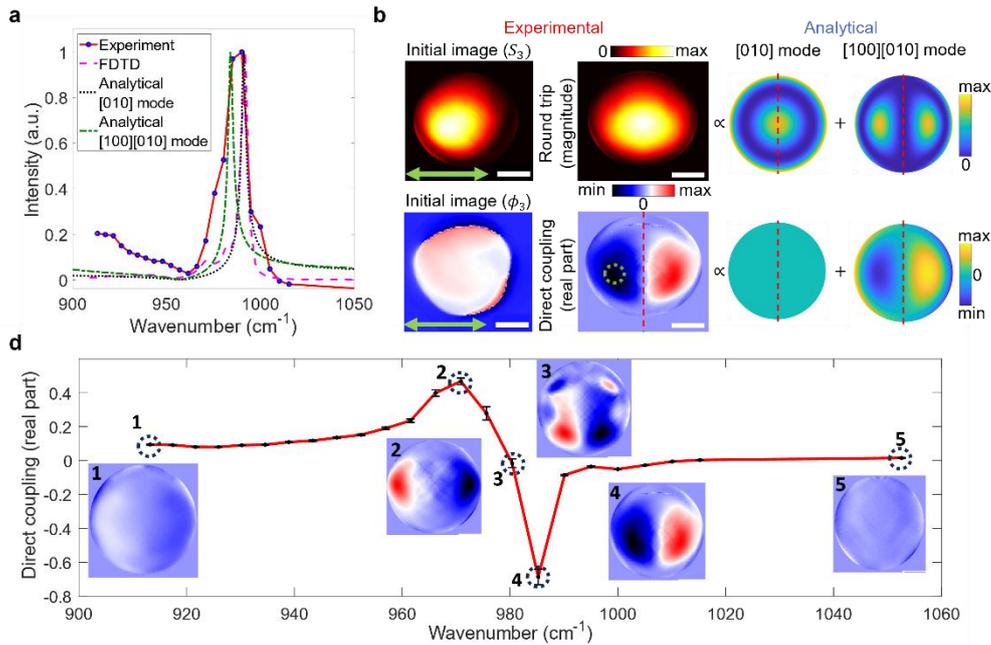

*Figure 3 – s-SNOM measurements and analysis of single α-MoO$_3$ nanoellipsoids (112/220 nm). a) Comparison between normalized absorption of the [010] dipole mode from FDTD simulation (purple dashed line), and analytical model (black dotted line), and the [100][010] higher order mode from the analytical model to experimental 3$^{rd}$ harmonic phase spectrum results from the nanoellipsoids. b) Analysis of the third harmonic s-SNOM signal at 985 cm$^{-1}$ (3$^{rd}$ harmonic) and separation into the complex representation of RTr and direct DCo along the symmetry axis (red dotted line). The green arrow indicates the direction of the incident beam polarization in the image plane. The measured data is compared to the combined prediction of RTr and DCo contributions of the [010] out-of-plane mode and the [100][010] in-plane mode from the analytical model. c) Spectral measurement of the direct coupling contribution for α-MoO$_3$ nanoellipsoids averaged over the green circular area in (a), showing resonance behavior. Snapshots of the direct coupling are shown in the graph. The snapshots show a flip after the resonance of the mode.*

The measured spectrum is compared to the normalized dipole mode resonance along the [010] crystal axis and the [100][010] higher-order mode resonance from the analytical model (Figure 2a). We also compare our result to the absorption spectrum from the Finite-difference Time-Domain method (FDTD, Lumerical Ansys) simulation of an α-MoO$_3$ nano-ellipsoids illuminated by a plane wave with an electrical field along the [010] crystal axis. The [010] mode peaks obtained from the analytical model and FDTD simulation show an excellent match with each other and with the location of the measured near-field peak at 988 cm$^{-1}$. The measured near-field amplitude spectrum of the nano-ellipsoids also shows evidence of a secondary resonance at ~970 cm$^{-1}$ in proximity to the dipole mode. Measurements on additional nano-ellipsoids confirmed this secondary resonance; see SI Figures S5 and S6.

The complex near-field signal that results from the interaction between the probe tip and single nano-ellipsoids resonant modes can be divided into two main components [33]: The first component is the Direct Coupling (DCo), which originates from the modes that are excited by the incident beam and then scattered to the far-field through coupling with the AFM tip. The DCo contribution image depends on the direction and polarization of the incident beam with respect to the direction of the excited mode. The second component is the Round-Trip (RTr) contribution, which involves the modes that are excited and subsequently scattered to the far-field by the AFM tip. The RTr contribution depends on the position of the tip relative to the sample during measurement. For ellipsoids-shaped nanoparticles, where the crystal axes align along the geometrical axis, the RTr contribution will exhibit an even symmetry according to the directional mode behavior. In DCo, the in-plane modes excited *via* the AFM tip have odd symmetry with respect to the crystal axis. The complex s-SNOM signal can thus be decomposed into these two contributions using their symmetry properties (see details in the Methods Section and Ref.[33]). We can calculate DCo by subtracting its replica mirrored across the odd symmetry axis from the complex near field image, and the RTr image can be generated by adding the mirrored signal. Note that, unlike isotropic structures, our modes are directional, and we adjusted the analysis accordingly.

Figure 3b shows the DCo and RTr contributions near the resonance peak in Figure 3a compared to the predicted DCo and RTr contributions from the out-of-plane [010] dipole and in-plane [100][010] high-order mode as predicted by the analytical model. The two left images show the 3$^{rd}$ harmonic near-field amplitude and phase image at 985 cm$^{-1}$. The RTr and DCo images extracted from the experiment are in excellent agreement with the combined RTt of the out-of-plane [010] dipole and in-plane [100][010] higher-order mode predicted by the analytical model, assuming a more dominant dipole mode contribution, for an additional explanation see Methods Section.

Figure 3c illustrates the normalized spectrum of the direct coupling, which is an average of measurements taken within the green-circled area in Figure 3b. Both the spectrum and accompanying images of the real part of the direct coupling exhibit resonance behavior, featuring a clear transition at approximately 980 cm$^{-1}$ with mode flipping after the resonance peak. Such behavior can be explained by considering an expression of the electric field close to a resonance of the form $^{36-38}$ $E \propto \frac{1}{\epsilon_z(\omega)-\epsilon_{zm}}|E_{l,m}\rangle$, which shows that there is a sign change when crossing a resonance.

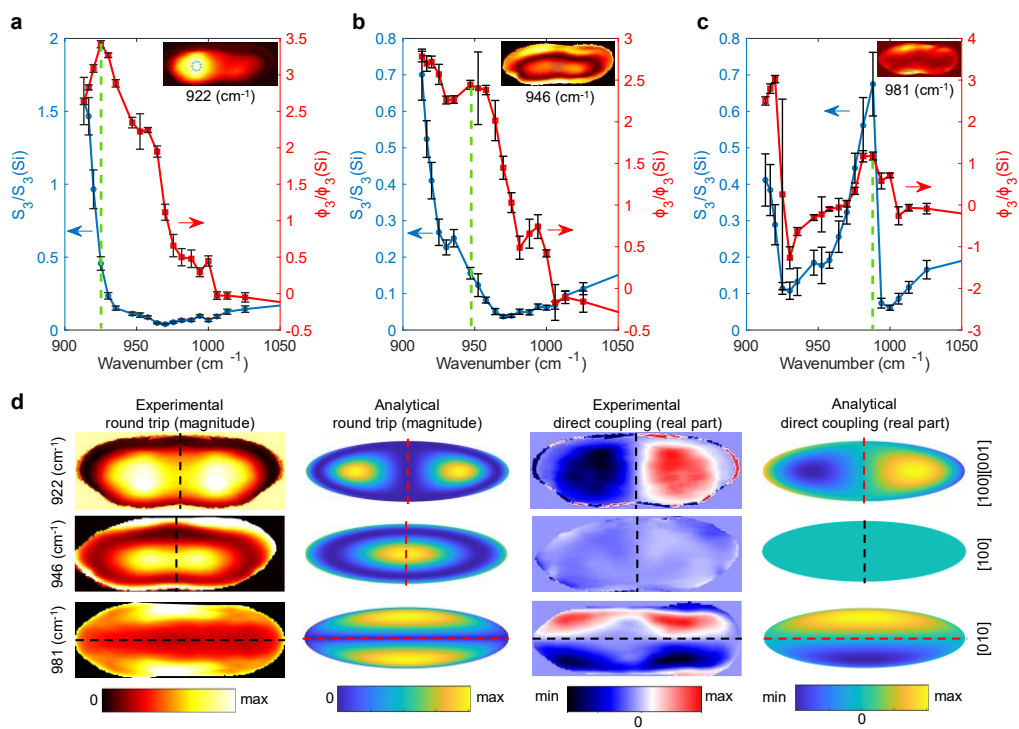

*Figure 4 – Mid-IR s-SNOM scan of α-MoO$_3$ nanorod (width/height 120 nm, length 400 nm) and results of the analytical model. a) Near-field amplitude and phase spectrum showing a resonance at 925 cm$^{-1}$. The insert shows a near-field amplitude image of the nanorod at 922 cm$^{-1}$ with a green circle indicating the measurement area for the spectrum. b) Near-field amplitude and phase spectrum showing a resonance at 946 cm$^{-1}$. The insert shows a near-field amplitude image of the nanorod at 946 cm$^{-1}$ with a green circle indicating the measurement area for the spectrum. c) Near-field amplitude and phase spectrum showing a resonance at 981 cm$^{-1}$. The insert shows a near-field amplitude image of the nanorod at 981 cm$^{-1}$ with a green circle indicating the measurement area for the spectrum. d) RTr and DCo for each measured resonance in a-c, and the corresponding prediction of RTr and DCo images calculated using the analytical model for the [100][001], [100], and [010] modes, respectively (see Figure 2f).*

Some synthesized α-MoO$_3$ nanoparticles have a rod-like shape, giving them higher aspect ratios than previously shown α-MoO$_3$ nano-ellipsoids. Figure 4 shows near-field measurements for a representative nanorod with a diameter of about 120 nm and a length of 400 nm (see SI Figure S7). Figure 4a-c presents the normalized spectra of the nanorod's phase and amplitude along the nanorod's radial focus, center, and edge, respectively. These spectra reveal three distinct resonances at approximately 925, 947, and 981 cm$^{-1}$, which correspond to a high order, out-of-plane dipole, and vertical dipole (in the short nanorod axis direction) directional modes, respectively. From the analytical model, we expect the dipole resonances along the [100] and [010] axes to occur at 940 cm$^{-1}$ and 988 cm$^{-1}$, respectively, and the high-order [100][001] mode peak to occur at 923 cm$^{-1}$, These values align very well with the experimentally measured resonances.

The dipole results also match the far-field FDTD simulation of the nanorod dipole modes (see SI Figure S8). Figure 4d compares the DCo and RTr images extracted from the experimental data near each resonance peak to the corresponding images from the analytical model. The experimental DCo and RTr distributions imaged at 922, 946, and 981 cm$^{-1}$ are in excellent agreement with the [100][001], [100], and [010] mode projections, respectively, predicted by the analytical model. This study showcases the remarkable capability of our α-MoO3 nanorods to enable both dipole and higher-order LHPhR across a wide range of MIR spectrum. Moreover, the notable variations in the spectral properties between the measured nano-ellipsoids and nanorods underscore the exceptional degree of resonance tuning achievable with our system.

**Conclusions**

α-MoO$_3$ biaxial nanoparticles were investigated for the first time, both theoretically and experimentally. These subwavelength nanoparticles, synthesized using a novel method of fs-laser pulse ablation in liquid, support two mid-IR LHPhR in the 8.5-11 μm spectral range. We develop a complete theoretical model of high-order resonances of general biaxial anisotropic nanoparticles in the quasistatic approximation. We obtained that the 1$^{st}$ and 2$^{nd}$ order analytical modes for biaxial anisotropic particles agree well with the experimental results. We observe 2nd order modes at frequencies even lower than the first-order mode's resonance due to the ellipsoid's geometry. Moreover, we find that in contrast to isotropic nanostructures, the resonances of high-order modes couple the permittivities in the different axes.

Our α-MoO$_3$ nano-ellipsoids exhibit sharp resonance in the MIR range, accompanied by extremely high Q factors. The high degree of confinement causes the LHPhR to undergo a blue shift of approximately 32 cm$^{-1}$, indicating the potential for broad tunability in the MIR range. Our nanorod-shaped α-MoO$_3$ nanoparticles demonstrate the ability to tune both RB2 and RB3 LHPhR within a single particle by modifying the aspect ratio, revealing a polarization-dependent resonator.

This research establishes that α-MoO$_3$ nanoparticles can support LHPhR throughout the MIR spectral range, offering high Q factors, polarization sensitivity, and tunability. These findings open new possibilities for IR detectors, image sensors, and photonic devices, emphasizing the importance of comprehensive understanding. Our results are expected to apply to other fields of physics such as quasi magnetostatics and heat conduction[42,43]

**Methods**

**α-MoO$_3$ nanoparticles synthesis**

MoO$_3$ pellets were used as ablation targets in the fs-PLAL process. The pellets were made from MoO$_3$ powder (Sigma-Aldrich) through a process of powder grinding, followed by palletization using a hard press at a pressure of 5 MPa. The resulting pellets were then sintered at a temperature of 560° C for 12 hours. XRD and Raman spectroscopy were used to analyze the pellets and confirm the formation of the α-MoO$_3$ phase.

The ablation process was carried out using a Ti: Sapphire femtosecond laser (Astrella, Coherent Inc., Santa Clara, CA, USA) with a central wavelength of 800 nm, a repetition rate of 1 kHz, a full-width at half-maximum of 500 fs, and an average power of 1.5 W. The laser beam was focused on the target using a converging lens with a focal length of 20 cm. A pre-prepared $MoO_3$ pellet was placed at the base of a beaker filled with 30 mL of acetone (Sigma-Aldrich). Acetone was chosen as the solvent because of its inherent stability, which helps to protect the pellet in its liquid environment. The presence of a liquid layer above the $MoO_3$ particle prevents exposure to ambient air. The Rayleigh range of the ablating femtosecond laser beam needs to be completely within the solvent layer above the sample for proper focusing of the beam. To minimize the potential impact of the laser on the targeted area, the pellet enclosed in the glass beaker was continuously moved using a translation stage. The translation stage can move in two dimensions and is operated by an automated program. The ablation procedures were conducted for 20 minutes. Afterward, the solution of nanoparticles was carefully stored in hermetically sealed containers. Subsequently, the solution was transferred to centrifuge tubes to facilitate debris separation. The centrifugation process was performed at a rotational speed of 5000 revolutions per minute (RPM) for 30 minutes. After centrifugation, unwanted debris precipitated in the lower region of the tube, while the top portion consisted exclusively of nanoparticles. A pipette was used to isolate these nanoparticles from any accompanying debris.

The resulting nanoparticle/acetone solution was further diluted by the addition of acetone. The samples were made by drop-casting onto a clean Si prime wafer and allowed to dry at room temperature.

Nanoparticle size and shape were measured using a combination of AFM for height measurement and Scanning electron microscopy for lateral dimensions (see SI).

**Near-Field Imaging:**

Near-field and AFM measurements were taken using a commercial s-SNOM (Neaspec Attocube). The system uses a Pt-coated AFM probe (PPP-NCLpt from NANOSENSORS) to scan the sample in tapping mode (160 kHz frequency 60 nm amplitude). A tunable Mid IR (910 – 1205 $cm^{-1}$) QCL (Mircat DRS daylight solutions) operating at 3 mW is used for optical near-field measurements. All results were taken from the 3$^{rd}$ demodulation order of the signal to reduce far-field noise. Furthermore, a pseudo-heterodyne measurement system is used to extract the optical phase and further reduce far-field noise in the measurements. Localized spectra were taken using point spectroscopy[44]. Consecutive scans of the sample were taken at different wavelengths and combined using a dedicated Python script[44,45] (correcting for the position of the particle and drift in each scan see SI for more details) to compose a high-resolution hyperspectral scan of the sample. The reported spectra are normalized with respect to the spectra of the Si substrate. Phase images were corrected for thermal drift of the interferometer reference arm using Gwyddion line scan correction algorithm.

The s-SNOM is used to measure the near-field signal of a resonating structure, it can be decomposed into four distinct scattering processes possible for a two-particle system. It is important to note that all these processes must involve coupling to the

probe; otherwise, they would be filtered out during demodulation. These contributions are: [33]

a) Signal arising from material contrast due to local polarizability, which is directly measured via the probe.

b) Round trip signal generated by resonant modes induced in the sample by the AFM tip, which then re-interact with the probe to scatter into the far field. As the coupling is exclusively through the AFM tip, both dark and bright modes can be excited regardless of their orientation to the incident beam polarization. These contributions should exhibit symmetry along the crystal axes of the material as they depend solely on the relative position of the AFM tip on the sample. Round-trip is a good representation of the sum of the magnitude of the complex image data.

c) Direct coupling is excited by the incident beam, which subsequently interacts with the probe. It is worth noting that this contribution also includes the opposite path of the tip exciting a resonant mode, which is then directly measured by the detector. This is included because, in reciprocal samples, they are identical. Direct coupling is a good representation of the z-component of the electric field of the resonant mode and as such is displayed as the real part of the complex image data.

d) Modulated scattering occurs when the incident beam is directly coupled in and out of the resonant structure but is perturbed by the presence of the probe near the structure, thus modulating the signal. Generally, this contribution is negligible for single-particle samples.[33]

Out of these four processes, only b and c have been shown to have significant contributions to the imaging of single nanoparticle modes [33].

Using the different symmetry properties of these contributions we decompose the measured field to the RTr and DCo contributions. The RTr contribution is always positive since it is proportional to $|E_z|^2$ [33]. The DCo contribution is proportional to $E_z$ so change the sign with the z-directional electric field of the mode. For the case of symmetrical nanoparticle geometries with a crystal axis aligned with the geometrical axis of the nanoparticle (such as nano-ellipsoids where the different crystal axis are aligned with the ellipsoid in different radial directions), we expect RTt to be symmetrical across the crystal axis and DCo to have the odd symmetry of the mode. Thus by using DCo odd symmetry properties, we can separate its contribution to the complex near-field signal $\sigma_n$ by subtracting its replica mirrored across the odd symmetry axis $\sigma_{n,\text{mirror}}$ to get $\sigma_{DCo} = \frac{1}{2}(\sigma_n - \sigma_{n,\text{mirror}})$, and the RTr image can be generated by adding the mirrored signal $\sigma_{RTr} = \frac{1}{2}(\sigma_n + \sigma_{n,\text{mirror}})$.

While the main excitation of high-order modes is from the tip, which behaves as a point dipole, the far field can excite the higher-order modes with lower intensity which can be approximated by $\left(\frac{2a}{\lambda}\right)^{n-1} \approx \left(\frac{1}{8}\right)^{n-1}$, where n is the mode order, and a is the particle size. At the same time, sharp interfaces between inclusions are known to generate scattered fields with short wavelengths and hence in our setup, the substrate-particle interface may result in the excitation of high-order modes. An additional mechanism to couple a far field to high-order modes is the coupling between the crystal modes $k \rightarrow k + mg$, where $g$ is the crystal wavevector, which decreases for large integer $m$, see Refs[38,46,47].

**FDTD modeling:**

The FDTD was performed using Lumerical software from Ansys. The sphere was modeled using dielectric data taken from [48]. The system was modeled as an ellipsoid on a semi-infinite Si substrate with a broadband Total-Field Scattered-Field (TFSF) source. The source is a pulsed plane wave propagating in the z-axis with polarization set along the x-axis. The crystal axis of the sphere is defined with respect to the direction of source polarization. Near-field images were taken from a monitor tangent to the sphere and perpendicular to the source propagation direction (in the x-y plane). This enables us to simulate the nanoparticle's far-field interaction and image the real component of the z-aligned electric field on the top of the nanoparticle. This field correlates the most with the direct coupling component of the near field. The simulation ran for 15000 fs to allow full decay of all residual fields. The mesh size inside the source was set to 2.5 nm.

The α-MoO$_3$ nano-ellipsoids were simulated as a $[r_x, r_y, r_z] = [110, 110, 55]$ nm ellipsoids with a crystal axis aligned as such: [010] – along the z-axis, [100] and [001] – are both pointed in the x-y plane.

The α-MoO$_3$ nanorod was simulated as an ellipsoid with a major axis at 200 nm and a minor at 60 nm. The crystal axis aligned as such: [001] – along the length of the ellipsoids, which matches previously reported growth of MoO$_3$ rod-like shaped particles[5,30].

**Acknowledgments**

The authors thank Dr. Zahava Barkay for the SEM analysis.